# Silicon carbide stacking-order-induced doping variation in epitaxial graphene

Davood Momeni Pakdehi [1]\*, Philip Schädlich [2], T. T. Nhung Nguyen [2], Alexei A. Zakharov [3], Stefan Wundrack [1], Florian Speck [2], Klaus Pierz [1]\*, Thomas Seyller [2], Christoph Tegenkamp [2], and Hans. W. Schumacher [1]

[1] Physikalisch-Technische Bundesanstalt, Bundesallee 100, 38116 Braunschweig, Germany
[2] Institut für Physik, Technische Universität Chemnitz, Reichenhainer Straße 70, 09126 Chemnitz, Germany
[3] MAXIV laboratory, 22484, Lund, Sweden

Email:      davood.momeni.pakdehi@ptb.de
            klaus.pierz@ptb.de

**Abstract**.

Generally, it is supposed that the Fermi level in epitaxial graphene is controlled by two effects: *p*-type polarization doping induced by the bulk of the hexagonal SiC(0001) substrate and overcompensation by donor-like states related to the buffer layer. In this work, we evidence that this effect is also related to the specific underlying SiC terrace. We fabricated a periodic sequence of non-identical SiC terraces, which are unambiguously attributed to specific SiC surface terminations. A clear correlation between the SiC termination and the electronic graphene properties is experimentally observed and confirmed by various complementary surface-sensitive methods. We attribute this correlation to a proximity effect of the SiC termination-dependent polarization doping on the overlying graphene layer. Our findings open a new approach for a nano-scale doping-engineering by self-patterning of epitaxial graphene and other 2D layers on dielectric polar substrates.

**Keywords**. Surface-dependent polarization doping, SiC spontaneous polarization, epitaxial graphene, hexagonal silicon carbide.

## 1. Introduction

The growth of epitaxial graphene on large-scale silicon carbide (SiC) substrates enables the fabrication of electronic devices for a wide range of technological applications in an industrial fabrication environment. [1–3] The role of SiC exceeds that of a simple carrier of the graphene sheet as it influences the basic electronic properties of the graphene in different ways and thus can be used for further modification of graphene's electronic properties. The most commonly used SiC substrates are of the hexagonal 4H and 6H polytypes, which exhibit a spontaneous polarization induced by the hexagonal stacking sequences in the SiC unit cell. The spontaneous polarization of the hexagonal SiC polytypes leads to a phenomenon called polarization doping, i.e., a *p*-type doping of the order of 6 - 9 × $10^{12}$ cm$^{-2}$ in so-called quasi free-standing graphene on hydrogen saturated SiC(0001). [4] On the other hand, epitaxial graphene residing on the buffer layer shows an *n*-type conductivity with a charge carrier density of the order of $10^{13}$ cm$^{-2}$. This is attributed to an overcompensation of the polarization doping by electron transfer from a donor-like buffer layer and interface states to the graphene layer. [4,5] Superimposed on these fundamental and spatially homogenous effects, the interplay between the hexagonal SiC and atop carbon layer is a source of various other intriguing phenomena, e.g., step-induced extrinsic resistance anisotropy in graphene [6–9], dislocation boundary domains [10], room-temperature strain-induced quantum Hall phase [11] and ballistic transport at SiC sidewalls [12], the Stark effect [13] and quantum photonics in SiC defect sites and color centers [14], as well as offering an excellent platform for growing other low-dimensional materials [15–17]. In general, the graphene properties on the SiC terraces are assumed to be uniform. These terraces, considering the stacking order in the unit cell of hexagonal SiC, result in inequivalent surface "terminations." Given that, some doubt is raised because theoretical investigations estimate a certain dependence of the doping on the stacking sequence and surface terminations of 4H- and 6H-SiC. [18–22] Moreover, a very recent nano-scale transport study reported up to a 270% variation (at low temperature) of the sheet resistivity for epitaxial monolayer graphene on two different terminated 6H-SiC terraces. [23]

In this paper, we experimentally show that the stacking terminations of the hexagonal SiC substrate have a biasing effect on the surface potential leading into a work function variation and thus a shift of the doping level of the overlying epitaxial monolayer graphene. Using the so-called polymer-assisted sublimation growth (PASG) technique, monolayer graphene on identical and non-identical SiC terraces was fabricated. In our study, various surface-sensitive measurement techniques, namely atomic force microscopy (AFM), scanning tunneling microscopy (STM), low-energy electron microscopy (LEEM), kelvin-probe force microscopy (KPFM), and X-ray photoemission electron microscopy (XPEEM) indicate different electronic properties of graphene on inequivalent SiC surface terraces types in association with their cubic and hexagonality nature. These SiC terraces can be clearly assigned to the specific stacking terminations within the framework of an extended SiC step retraction model.

## 2. AFM analysis of graphene and buffer layers on identical and non-identical 6H-SiC terraces

The epitaxial buffer and graphene layers in this work were grown on semi-insulating 6H-SiC samples with a nominal miscut of about $-0.06°$ toward $[1\bar{1}00]$ (from II−VI Inc.). Epitaxial growth was carried out in a horizontal inductively heated furnace. [24] The buffer layer and graphene samples were grown by the PASG technique in an argon atmosphere (~900 mbar) at 1400 °C and 1750 °C, respectively. [8,9,25] The control of the surface morphology was attained by taking into account the influence of Ar flux during the sublimation growth. [9]

**Figure 1**(a-d) and (g-j) show AFM images of the epitaxial monolayer graphene and buffer layer, respectively, with two types of surface morphologies. The origin of the different surface morphology will be explained in the next section. The AFM topography images of the graphene samples in **Figure 1**a and b show that one sample exhibits regular ~0.75 nm step heights while the other one displays a step pattern consisting of alternating ~0.25 nm and ~0.5 nm high steps. The phase images of these samples in **Figure 1**c and d show a very interesting behavior. For the ~0.75 nm stepped surfaces, the same phase is observed on all terraces. Only step edges appear as narrow regions with increased phase. On the other hand, ~0.25/~0.5 nm stepped surface clearly shows an alternating phase (**Figure 1**d), which changes from one terrace to the next. As for the other sample, step edges appear as narrow regions with increased phase. The observation of two different phase values indicates different material properties of the graphene layer on neighboring SiC terraces. Importantly, this phase contrast is not caused by a different number of graphene layers. The integrated Raman spectra (areal scan over $20 \times 20$ µm$^2$) in **Figure 1**e and f reveal a similar spectrum for both samples and a typical 2D-peak (at ~2724 cm$^{-1}$ and full-width-half-maximum of about 33 cm$^{-1}$) which proves that both samples are uniformly covered with monolayer graphene. [9,26,27] Similar contrast was also seen in the scanning electron microscopy of the sample with sequential terrace-steps. [28]

The AFM investigation of the buffer layer samples (**Figure 1**g-j) leads to a very similar result. Again, the regular ~0.75 nm stepped SiC terraces (see **Figure 1**i) show no phase contrast and only for the binary ~0.25/~0.5 nm stepped terraces an alternating AFM phase contrast is observed (see **Figure 1**j). The integrated Raman spectra in **Figure 1**k and l show a broad buffer layer related vibrational band, which indicates a homogenous buffer layer coverage. [29]

The agreement of the phase contrast of graphene and the buffer layer samples for the same substrate step structure clearly points to a substrate related effect changing the properties of the overlying layer, be it the epitaxial graphene or the buffer layer. An intrinsic effect of the graphene layer itself thus can be ruled out. In AFM experiments, the phase image contrast arises from local variations of the energy dissipation in the tip-surface interaction, which results in damping and shift of the tip's oscillation frequency giving information about the chemical/mechanical/electrical heterogeneity of a surface. [30] Thus, the observed phase contrast on non-identical terraces is associated with a change of the surface properties originating from the different SiC stacking terminations of the corresponding terraces below. Therefore, in the following, the formation and nature of the SiC surface terraces are analyzed.

## 3. Step retraction model of 6H-SiC(0001)

The formation of the SiC terraces during graphene growth can be understood in the framework of the SiC step retraction model applied to the 6H polytype SiC(0001) substrate. **Figure 2**a illustrates the unit cell of 6H-SiC consisting of 6 Si−C bilayers called A, B, C, A, C, B (from bottom to top). For our investigation, we focus on the 6 resulting Si-terminated surface-terraces which differ in three inequivalent stacking sequences of the underlying Si-C bilayers, i.e., S1, S2 and S3, and another three stacking sequences, S1*, S2*, S3*, which are equivalent to the first ones but rotated by 60°, see **Figure 3**b. [31] The number gives the number of SiC bilayers between the surface and the first hexagonal stacking arrangement. For simplicity, we name them S1, S2, and S3 if the rotation can be neglected. The eclipsed and staggered orientation of subsequent Si−C tetrahedra are called hexagonal (*h*) and cubic (*k*), respectively, which leads to discrete *h, k,* and *k* stacking orders of A, B, and C. In a more detailed model, one can assign to each atomic layer a hexagonal or cubic orientation, which is sketched in **Figure 2**a. [14] For the on-bonds in axial configuration ([0001] direction), this results in *hh*, *kk,* and *kk* stacking for the layers A, B, and C, respectively. The off-bonds (basal configuration) are described by *kk* (for S2) but also by mixed *hk* (for S3) and *kh* (for S1) stacking orders. The hexagonality of each surface terrace can be considered as the joint hexagonality of the corresponding on- and off-bonds.

The schematic representation of the terrace structure with single Si−C steps (1L) of ∼0.25 nm, in **Figure 2**b and c, can be regarded as the initial surface of the low-miscut angle substrates used here. During graphene growth at high temperatures, a restructuring of the SiC surface takes place which can be described by retraction of individual Si–C bilayers with different velocity in a so-called step retraction model. [32–34] The step retraction is driven by the minimization of the surface energies, which depend on the surface hexagonality of the individual terraces. [22,34] This step retraction velocity is indicated by the length of the horizontal arrows in **Figure 2**b. The high retraction velocity of the S1 surface can be attributed to the strongest hexagonality (*hh* on-bond) of this layer in agreement with refs. [32,33]. Thus, the corresponding S1 surface disappears at first. S2 and S3 remain, which results in a periodically stepped surfaces with ∼0.25 nm (1L) and ∼0.5 nm (2L) step-heights, as observed in the AFM image in **Figure 1**b and h. This situation is sketched as an intermediate state in **Figure 2**c. From this model, we can clearly assign the S3 terrace being above a 1L step and an S2 terrace above a 2L step. It is further assumed that S2 is the most stable surface because of its least hexagonality (*kk* on-bond and *kk* off-bond) and, therefore, the width of the S3 terrace (*kk* on-bond but *hk* off-bond) is decreasing faster than S2, which agrees with the model in ref. [33,34] but not ref. [32]. The width of the initially wider S3 terrace (see buffer layer AFM image in **Figure 1**h and ref. [9]) decreases, and for advanced step retraction the S3 terraces become narrower than S2. This situation is found for all graphene samples, see **Figure 1**b and in refs. [8,9,25]. (Such a pattern could not be primarily formed if S2 would retract faster than S3.) Finally, only the stable S2 terraces remain with step heights of ~0.75 nm (3L), sketched as the final state in **Figure 2**c. This situation is observed in the AFM images in **Figure 1**a and g.

The preparation of a ~0.25/~0.5 nm stepped surface with alternating S2 and S3 terraces of nearly 100% efficiency is a specific advantage of the PASG method. [8] The fast formation of the buffer layer by the additional polymer-related carbon supply stabilizes the SiC surface and reduces the step bunching velocity. As a result, the terrace structure can be "frozen in" at the intermediate state with S2 and S3 terraces before the final state with S2 surfaces only, is reached. [25,28]

## 4. Assignment of the SiC terminations below the graphene layer

A bright-field (BF) LEEM image of the ~0.25/~0.5 nm (1L/2 L) stepped graphene sample is displayed in **Figure 3**a. In the upper part of this image, a regular contrast pattern of alternating narrower brighter and wider darker stripes is observed. From the similarity to the AFM pattern, the BF-LEEM stripes are identified as the terraces S2 (wide) and S3 (narrow). Note that the step edges between the terraces lead to the thin dark lines in **Figure 3**a. From the correlation between terrace width and step height (narrower S3 being above a 1L and wider S2 being above 2L step), we can ascribe the corresponding step-heights to the boundaries between areas of different brightness, which is visualized by the top blue profile in **Figure 3**a. The underlying regular SiC step structure is sketched in part (i) of **Figure 3**e (For clarity the covalently bonded buffer layer and the overlying graphene layer are left out in this sketch). The bright-field image depicts the reflected intensity of the $0^{th}$ order low-energy electron diffraction (LEED) spot. The usual attribution of the BF contrast to a graphene thickness variation [35,36] is not valid here since the sample is unambiguously covered only with monolayer graphene. As will be discussed further below, the different brightness in the BF-LEEM images can be related to a variation of the surface potential, which is induced by the stacking of the SiC crystal underneath.

Next to the very regular S2/S3 terrace pattern in the upper part of this BF-LEEM image, there also irregular terrace configurations are observed in the lower part of **Figure 3**a. As an example, the line (ii) crosses a bright/bright transition, and line (iii) include a dark/dark transition. The corresponding step structure is sketched in the parts (ii) and (iii) in **Figure 3**e and can be explained by a ~0.75 nm (3L) terrace step in both cases. Such terrace configurations occur since the terraces of the starting SiC substrate are not entirely equal in length (and width). On both sides of the ~0.75 nm steps, the same terrace type (both are either S2 or S3) is found, which results in the same LEED reflection intensity, i.e., no contrast in the BF-LEEM image is observed. This interpretation allows a consistent explanation of the bright and dark areas of the complete BF-LEEM image by the corresponding terraces types (i), (ii), and (iii), as marked in **Figure 3**e.

For a test of the surface structure model, also the dark field (DF) LEEM image was recorded using the moiré diffraction spot marked in **Figure 3**d. The moiré diffraction spots arise from multiple scattering at the graphene and SiC lattices and thus carry information about the SiC lattice orientation. The resulting DF-LEEM image is shown in **Figure 3**b. It also exhibits a sequential binary contrast pattern, but interestingly, the width of both stripes is wider and not congruent with the stripes observed in the corresponding BF image, **Figure 3**a. For the regular stepped upper part of the LEEM images (see the line (i) in **Figure 3**a and b), we can deduce that two terraces (one bright and one dark stripe in BF) comprise one stripe in the DF image. This situation is sketched in **Figure 3**c. The origin of the DF contrast is the 60° rotation between the equivalent surface terminations Sn/Sn*, which causes a 60° rotation of the respective threefold diffraction spots patterns (see **Figure 3**d) of the terrace, which is sketched as green and purple areas in **Figure 3**b. Thus, when crossing a step from S to an S* (or vice versa) terminated terrace (S2-S3*, S2*-S3, S2*-S2, S3*-S3) the LEED pattern is rotated by 60° and a change of the DF contrast appears. On the other hand, no contrast inversion takes place when crossing a step from S3 to S2 or from S3* to S2*, since the direction of the underlying SiC is preserved. Using this interpretation also the irregular regions of the LEEM image can be explained, which is sketched in **Figure 3**e for the lines (ii) and (iii). With the described model, the different contrast patterns of the dark- and bright-field LEEM images complement each other and result in a consistent view of the underlying SiC terrace structure. With the input of the AFM results and the presented step retraction model, an unambiguous assignment of the step heights is possible, as are labeled in **Figure 3**a-b.

## 5. Verification of the SiC terrace model by scanning tunneling microscopy

A confirmation of the presented surface terrace model and a visualization of the SiC surface crystal orientation is provided by a low-temperature scanning tunneling microscopy (STM) investigation of the graphene sample. **Figure 4**a shows a large-scale topography of the irregular stepped area (iii) of **Figure 3**a. From the measured step heights (3L, 2L, and 1L), an assignment of the underlying SiC surfaces next to each terrace step is possible. For the applied STM energies, the graphene appears transparent and a high-resolution image of the buffer layer structure is resolved with the characteristic ($6\sqrt{3} \times 6\sqrt{3}$)R30° superlattice. [37,38] High-resolution images of the areas in the vicinity of the three consecutive terrace steps are shown in **Figure 4**b-d. They show a smooth buffer layer formation.

On all four terraces, S2*, S2, S3*, and S2*, see **Figure 4**b-d triangular-shaped structures are identified, partly marked as red/yellow triangles. Such corrugation structures are known to form during growth and surface reconstruction. [39–42] A close inspection of the STM images in **Figure 4**b-d reveals that the orientation of the triangles above and below a terrace edge is different for the three step heights. (For clarity the orientation and the rotation angle are depicted above each image.) For the consecutive terraces S2* to S2 (3L step of ~0.75nm) in **Figure 4**b and S2 to S3* (2L step of ~0.5 nm) in **Figure 4**c the triangle orientation rotates by 60°. The same rotation is observed for the ($6 \times 6$) quasi corrugation (diamonds sketched in blue) and the ($6\sqrt{3} \times 6\sqrt{3}$)R30° diamonds (sketched in yellow) of the buffer layer. Only for the 1L (~0.25 nm) step crossing from S3* to S2* terrace the triangle orientation does not change as well as the direction of the diamonds. Note, the 120° (or −60°) rotation of the diamonds of the first and the final S2* terrace, in **Figure 4**b and c, is a result of the 3-fold rotational symmetry. Both terraces are otherwise equivalent. The observed rotations of the SiC surface lattice are in excellent agreement with the SiC terrace model depicted in the corresponding area (iii) of **Figure 3**e and it is consistent with the step-retraction model, the μ-LEED, and DF-LEEM results, which showed that for the S to S* crossings the underlying SiC surfaces are accompanied by a 60° rotation, but the crystal orientation is retained for all S* to S* or S to S crossings. The tight correlation of the ($6 \times 6$) and the ($6\sqrt{3} \times 6\sqrt{3}$)R30° diamond rotations across the terrace steps show very instructively that the buffer layer strictly follows the rotation of the underlying SiC lattice.

## 6. Surface potential and work function measurements of graphene on non-identical SiC terminations

So far, AFM phase images reveal that the surface properties of graphene monolayers are different depending on the stacking termination of the underlying SiC terrace, which is in good agreement with the observed reflectivity contrast in the BF-LEEM images. In this section, we use additional methods to quantify the energy difference between the terraces and we discuss possible origins.

First, ambient Kelvin probe force microscopy in amplitude mode (KPFM-AM) is used to measure the surface potential ($\Delta V$) and the work function $\phi$ of the graphene layer. [43,44] The AFM topography of a binary 1L/2L stepped monolayer graphene sample is shown in **Figure 5**a. The measured step heights allow the assignment of the terraces to the underlying SiC surfaces S2 and S3. The KPFM surface potential maps of the same area are displayed in **Figure 5**b and a section enlargement in **Figure 5**d. The weak binary contrast pattern (dark and light grey) of neighboring terraces is very similar to the AFM phase image in **Figure 1**d. The potential values of two neighboring terraces are calculated by taking the median values from an area of $100 \times 600$ nm$^2$ (dashed rectangles in **Figure 5**d). The corresponding histogram of the potential values in **Figure 5**e clearly shows the difference between the S2 and S3 surfaces. The potential difference of monolayer graphene on both terraces S2 and S3 results in $V_{S2} - V_{S3} = 9 \pm 2$ mV (in air) which corresponds to a work function difference of about $\Delta\phi_{S2-S3} \approx -10$ mV. ($\Delta\phi_{S2-S3} = (\phi_{Probe} - eV_{S2}) - (\phi_{Probe} - eV_{S3})$). [36,45]

The KPFM map in **Figure 5**b also shows the elevated surface potential of bilayer graphene (BLG) spots (red areas), which have formed around a substrate defect on an S3 terrace. The homogenous potential of the bilayer graphene is also used to measure the relative difference to the monolayer graphene on both types of terraces. The local potential variation along the line scans correlated with an S2 and S3 terrace (magenta and green line in **Figure 5**b) is measured and the values are plotted in the histogram in **Figure 5**c. A clear difference between the S2 and the S3 related potential is observed with a value of $V_{S2} - V_{S3} = 12 \pm 2$ mV in reasonable agreement with the previous value from areal integration. The potential difference between monolayer and bilayer graphene is much larger with a value of about 60 mV, which is consistent with literature data but smaller than the reported values for vacuum measurements. [43–47] It is known that the absolute surface potential value is reduced by moisture and atmospheric adsorbates on the surface as well as intrinsic limitations of AM-KPFM and thus makes it difficult to precisely assess the surface potential difference. [44,45,48,49] It is worthwhile to mention that a potential difference of 10–20 meV for monolayer graphene on different terraces can also be seen in AM-KPFM (in air) measurement in ref. [43], which was not discussed.

As a second approach, low-energy electron reflectivity (LEEM-IV) measurements are used to deduce the graphene specific energy from the energy dispersion of the reflected electron beam in BF µ-LEED geometry. [50] The LEEM-IV measurements in **Figure 6**a were performed in vacuum with thermally cleaned surfaces on two neighboring terraces, S2 and S3. All LEEM-IV spectra show one prominent minimum, which is the signature of monolayer graphene in agreement with the Raman measurements in **Figure 1**f. [51,52] **Figure 6**b displays a map of the position of the minimum of the LEEM-IV spectra taken for each pixel of the BF-LEEM image in **Figure 3**a. The lateral distribution of the minimum energies in the LEEM-IV map in **Figure 6**b shows an identical contrast pattern (compare the upper part of **Figure 3**a). This shows that next to the reflected intensity of BF-LEEM, also the energy of the minimum is correlated with the underlying SiC surface termination.

For the neighboring terraces S2 and S3, an energy difference of ($E_{S3} - E_{S2}$) of (60 $\pm$10) meV is estimated from the minimum energy histogram in the inset of **Figure 6**b, which indicates a distinct difference in the graphene properties on both terraces. It worthwhile to mention that nearly the same value was measured for the graphene on 4H-SiC with non-identical terraces, however this polytype does not show a sequential pattern of terraces like 6H-SiC. [8] Although, in an early publication, the minimum energy was related to the graphene work function under the assumption that the reflectivity spectrum is related to discrete energy levels in the conduction band along the $\Gamma$-$A$ direction of the graphene band structure, [52,53] a recent study suggests, that the graphene interlayer bands play the decisive role. Following this model, the minimum position depends on the distance between the buffer and graphene layer and the corresponding correlation-exchange potential. [50,54] To explain the systematic shift of the LEEM-IV minima, at least one or both parameters should be different for graphene on S2 and S3 terraces. Since High-resolution STM measurements [23] on comparable 1L/2L stepped PASG graphene samples revealed fluctuating step heights variations (with smaller as well as larger on both 1L and 2L steps), thus systematic step heights differences can be excluded. This suggests a considerable difference in the correlation-exchange potential for both terraces, which is sensitive to the different charge densities. More detailed model calculations are necessary to verify this assumption.

As the third complementary method, X-ray photoemission electron microscopy (XPEEM) was applied, which directly visualizes the work function variation on the surface. [55] From the broad secondary electron energy distribution generated by X-ray excitation (photon energy of $h\upsilon$ = 80 eV), those with a low kinetic energy of 1 eV are selected and measured. Thus, the intensity map visualizes spatial variations of the work function. The XPEEM image of a 1L/2L stepped graphene sample is displayed in **Figure 6**d, and it shows the typical contrast pattern of the alternating surface terraces: narrow (S3) and wider stripes (S2). The XPEEM image taken at the same position as the BF-LEEM image (**Figure 6**c) shows a congruent pattern which verifies this assignment. In contrast to BF-LEEM images where an arbitrary contrast is obtained, the intensity in the XPEEM image is unambiguously correlated with the magnitude of the work function, namely, the layer with the lower work function generates a brighter contrast. [55] Thus, the S2 related graphene terraces (light grey stripes) have a lower work function compared to S3 terraces (dark grey stripes) in agreement with the KPFM result. This work function difference and the related potential difference are regarded as the reason for the contrasts observed in the AFM phase and BF-LEEM images. The different interaction of the AFM tip and the varying potential on the non-identical terraces results in a damping and a phase shift of the tip's frequency. In the BF-LEEM experiments, the slightly different surface potential and charge state, respectively, varies the reflection behavior on non-identical terraces and results in a variation of the reflected intensity of the $0^{th}$ order electron beam.

KPFM and XPEEM reveal in good agreement a lower work function for monolayer graphene on S2 compared to S3 terraces, and also the LEEM-IV result points to this direction. A reliable value for energy difference of ~10 meV was measured by KPFM, knowing that this value might depend on the environmental conditions. [49] The strong XPEEM contrast points to a higher energy difference, but beam damage effects prevent a more precise estimate of the energy difference. More detailed studies are necessary to clarify the absolute value of the energy difference.

Up to now, other experimental results are lacking. The local measurements of the sheet resistivity by high-resolution scanning-tunneling potentiometry (STP) on similar 1L/2L stepped samples (thermally cleaned) have indicated a difference of the average graphene sheet resistance by ~14% at room temperature and ~270% at low temperature (8 K) on both terraces S2 and S3. [23] For terraces connected by a 3L step, a smaller variation of < 3% is measured. [23] The analogy to the results presented here supports the idea of a strong impact of the SiC terrace termination on the graphene properties.

The presented measurements convincingly show a correlation of the SiC substrate termination with the electronic properties and the work function of epitaxial monolayer graphene. The graphene's $n$-type doping level is mainly determined by two effects, namely, an overcompensation of the SiC bulk polarization doping by donor-like buffer layer and interface states. [4,5] The strong intrinsic bulk polarization in hexagonal SiC substrates produce a negative pseudo-charge at the SiC surface, which induces positive charges in the graphene to account for overall neutrality. This polarization doping effect shifts the Fermi energy in freestanding monolayer graphene far below the charge neutrality point to $E_D - E_F = -0.3$ eV. [4] A specific terrace related polarization effect can be deduced from ab-initio pseudo charge calculations, which show a different valence band charge density for cubic and hexagonal SiC layers in the 6H polytype and a different charge density depending on the distance to the next underlying hexagonal layer. [20,21] In a recent publication, the total polarization doping effect of SiC surface and bulk was investigated by standard density functional theory calculations which show a SiC surface termination dependent doping variation of $2 \times 10^{12}$ cm$^{-2}$ for free-standing monolayer graphene. [18] However, the corresponding shift of the Dirac point of ~100 meV is larger than the KPFM value. For the exact modeling of the investigated epitaxial graphene samples, the impact of the intermediate buffer layer and interface states must be taken into account. A variation of the graphene work function induced only by a different distribution of buffer layer related donor-like states is excluded by the observation of AFM and BF-LEEM contrast patterns in buffer-layer free, $p$-type H-intercalated, quasi-freestanding monolayer [9] and also quasi-freestanding bilayer graphene. [to be published] The observation of an AFM phase contrast of the insulating buffer layer sample, see **Figure 1**j indicates that the SiC surface-related polarization effect also exists in the absence of graphene and the corresponding donor-like states, therefore it bears no relation to an interplay between the buffer layer and graphene.

This discussion shows that the assumption of a SiC stacking termination dependent polarization doping can explain the presented experimental results. Other effects, e.g., an influence of terrace dependent stress is less probable by the observation of strong contrast patterns also in more relaxed free-standing mono and bilayer graphene. [9] [to be published] A partial influence of different defects at both SiC terraces types can, however, not be ruled out. An example of a possible defect state in hexagonal SiC is the basal vacancy/divacancy, which can occur in a quasi-cubic or quasi-hexagonal position (shown in **Figure 2**a) with an energy difference of 0.03 eV. [56] Both quasi-positions are typical for the cubic S2 (0% hexagonality) and S3 (50% hexagonality) surfaces. To which extend such an effect plays a role requires further studies.

## 7. Summary and conclusions

In summary, by using a variety of measurement techniques (AFM, STM, µ-LEED, LEEM, LEEM-IV, KPFM, XPEEM) we have for the first time evidenced a direct dependence of electronic properties of epitaxial graphene on the underlying SiC stacking termination. This was realized by employing advanced epitaxial growth techniques including the PASG method. A periodic sequence of two different SiC terraces with the distinction in cubic and hexagonal nature of the surfaces were prepared, which develop during the high-temperature graphene synthesis. The terraces in the stacking orders are unambiguously identified as S2 and S3, as scrutinized in AFM and BF-LEEM and supported by additional DF-LEEM and STM measurements. The formation of the observed terraces is successfully interpreted in a SiC step retraction model in which the step retraction velocity increases with the hexagonality of the SiC surface layer. The KPFM and XPEEM results explicitly indicate an alternating work function of the graphene on periodic SiC surface terraces, which confirms for the first time a theoretical prediction in which the graphene doping depends not only on the bulk polarization but also on a SiC termination dependent polarization doping effect. A value of about 10 meV was estimated for the work function difference of monolayer graphene on S2 and S3 terraces from KPFM measurements. The periodically modulated surface potential self-ordered by the underlying SiC terraces could act as a template for further graphene functionalization schemes on sub-microscale structures. Moreover, our findings are applicable to other polar dielectric substrates and other sub-dimensional systems. For graphene on 4H-SiC(0001) substrates with only two different starting surfaces, we found no periodical variation of the surface work function, since step retraction results in the formation of equivalent S2 and S2* terrace terminations.

## 8. Acknowledgments

We gratefully acknowledge M. Wenderoth and A. Sinterhauf for valuable discussions. This work was supported in part by the Joint Research Project "GIQS" (18SIB07). This project received funding from the European Metrology Programme for Innovation and Research (EMPIR) co-financed by the Participating States and from the European Unions' Horizon 2020 research and innovation programme. A. A. Z. acknowledges the support from Stiftelsen för Strategisk Forskning (project RMA15-0024). We thank the support DFG Project Te386/12-1. We also acknowledge funding from the European Commission Graphene Flagship Core2 (grant agreement 785219).

## 9. Experimental section

**Sample preparation**. Samples in this study were grown on semi-insulating 6H-SiC(0001) specimens (5 × 10 mm$^2$) with a nominal miscut of about −0.06° toward [1$\overline{1}$00]. The SiC samples were first cleaned with acetone/ isopropanol in an ultrasonic-bath (USB) (15 min, 40 °C). The PASG method was applied using a so-called liquid-phase deposition (LPD) in a way that the samples were immersed into a mixture of AZ5214E polymer with isopropanol (25ml/ 5ml) and introduced to the USB (15 min, 40 °C). The specimens were subsequently rinsed with isopropanol (30 sec) before being spin-dried. [25,28] This results in nano-sized polymer adsorbates on the sample helping the growth uniformity. The buffer layer and graphene samples were grown in an argon atmosphere (~900 mbar) at 1400 °C and 1750 °C, respectively. [8,9,25] The control of the surface morphology was attained by several optimizations of growth parameters, including the influence of Ar flow-rate during the sublimation growth. [9]

**Raman spectroscopy**. Confocal micro-Raman mappings were performed at ambient conditions with a LabRAM Aramis Raman spectrometer (Horiba) equipped with a 600 mm$^{-1}$ grooves holographic grating, a frequency-doubled Nd: yttrium-aluminum-garnet laser emitting at 532 nm ($E_L$ = 2.33 eV), and a 100× (NA 0.9) objective to focus the excitation laser onto the surface of the samples. Surface Raman mapping images were captured over (20 × 20) μm$^2$ areas in backscattering mode using a piezo-driven XY-stage (PI) and a scanning step size of 0.1 μm.

**Atomic force microscopy/ kelvin probe force microscopy.** The AFM investigations were performed using two AFM setups: (i) the NANOStation AFM equipped with the SSS-NCLR silicon tips fabricated by "Nanosensors™," and (ii) the NanoWizard AFM system made by JPK instruments AG. The latter also enables KPFM (AM-mode) investigations. For this, conductive coated (Cr/Pt, 5 nm/25 nm) micromachined monolithic silicon tips (Tap300E-G, r=25nm) from "budget sensors" was supplied. In the AM-KPFM, the mapping of the topography and surface potential of the sample was performed in two-pass mode, in which first in tracing-pass the AFM in tapping mode collects the topographical data over the measuring profile line and this data is then used in retracing-pass at a set ascended height above the surface (<5 nm) to map the surface potential.

**Scanning tunneling microscopy.** The atomic structures were investigated using an Omicron low-temperature STM system at 77 K with a tungsten tip. Before characterization, the sample was degassed in situ at 500 °C for two hours.

**Low-energy electron microscopy/ X-ray photoemission electron microscopy**. There two sets of LEEM measurements were presented. The first (**Figure 3** and **Figure 6**b) was performed using a LEEM (FE-LEEM P90, Specs). The diffraction spots for dark-field imaging were selected by means of an aperture in the back focal plane of the objective lens. The second LEEM (**Figure 6**a and **c**) and XPEEM (**Figure 6**d) measurements were carried out at MAXPEEM beamline at 1.5GeV ring of MAXIV synchrotron. The beamline houses an AC LEEM microscope from Elmitec. Prior to each characterization, the samples were degassed in situ at 500 °C for 2 hours.

## 11. Figures

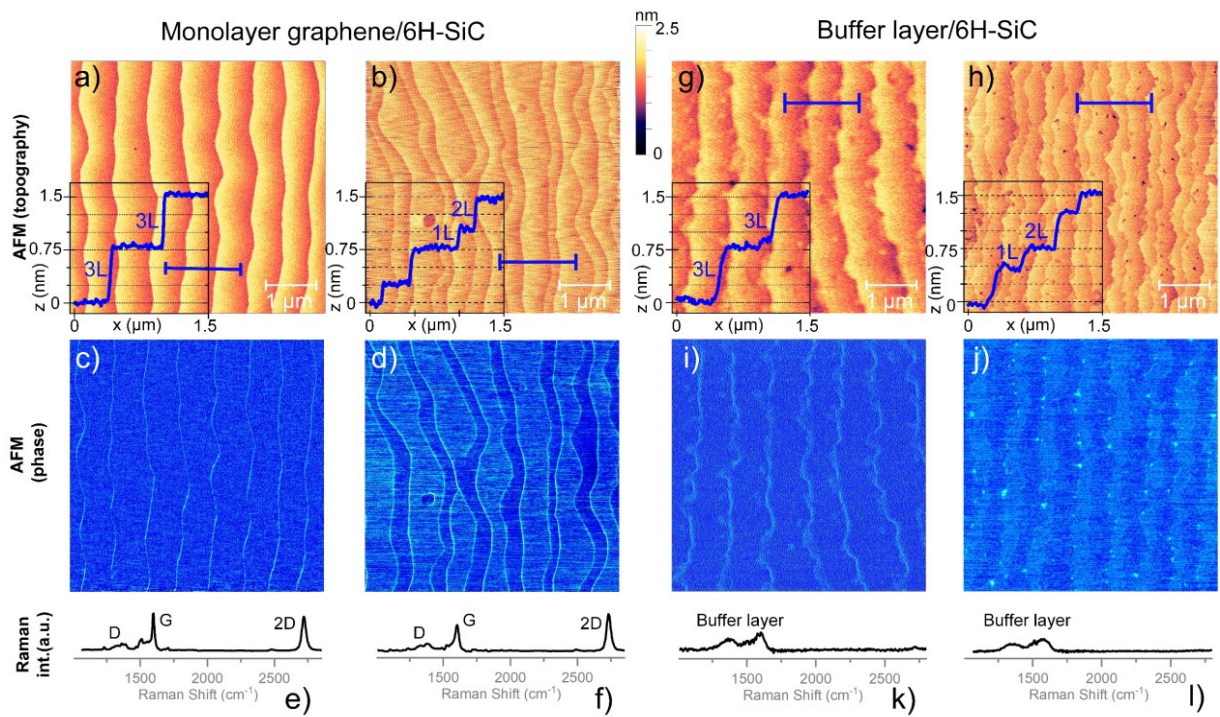

## Figure 1.

Atomic force microscopy (AFM) images of (a-d) epitaxial monolayer graphene and (g-j) buffer layer on 6H-SiC substrates with different terrace step heights of ~0.75 nm and sequential pairs of ~0.25/~0.5 nm, respectively. The cross-sections in the inset of (a, b, g, h) are taken along the blue line. (c, i) The phase images of the homogeneously stepped (~0.75 nm) samples show no phase contrast except for an increased phase at the step edges. (d, j) Only the phase images of the samples with step pairs (~0.25/~0.5 nm) show a sequential contrast on the terraces. (e, f, k, i) Raman spectrum of each sample. The displayed spectra were integrated over 14000 single spectra from an area of 20 µm × 20 µm. (e, f) The narrow 2D line widths of around 30 cm$^{-1}$ indicate that the graphene sample is thoroughly covered with monolayer graphene. (k, l) The broad vibrational density of states distribution indicates the existence of buffer layer graphene.

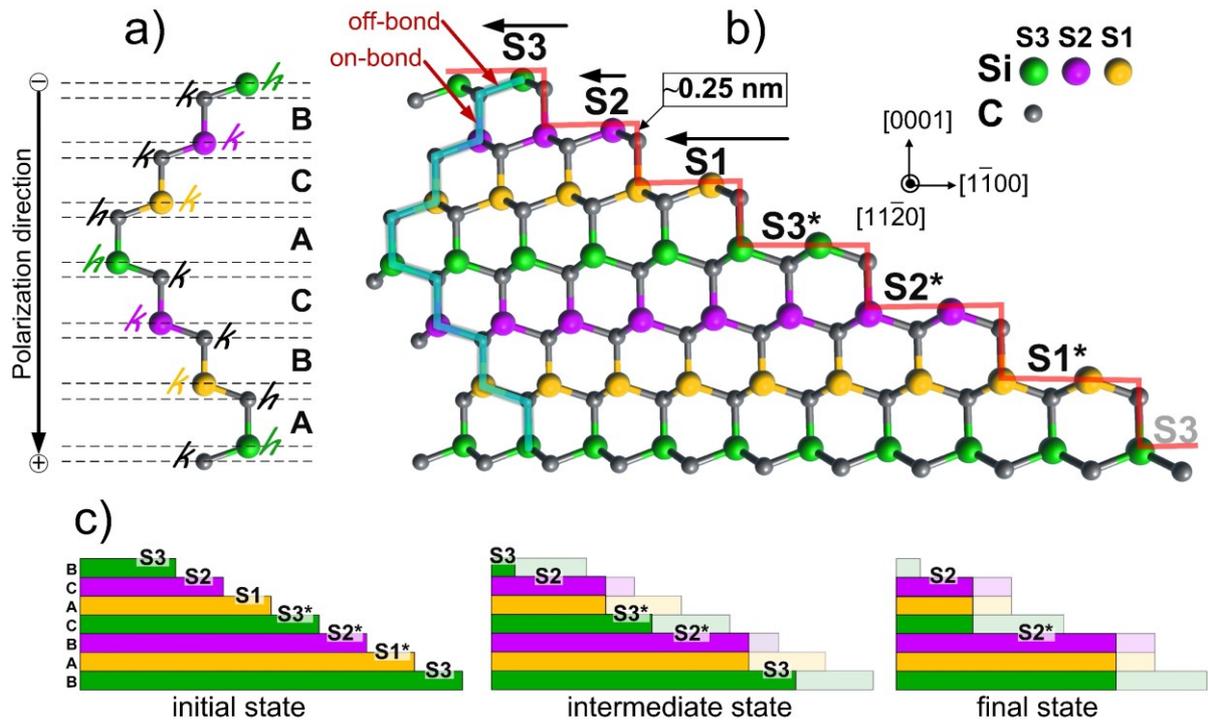

**Figure 2.**

Structural model of the 6H-SiC(0001) substrate and schematics of the corresponding step patterns in the step retraction model.

(a) Layer sequence of the Si-C bilayers in the 6H unit cell denoted as ABCACB. Each atomic layer is denoted by its atomic stacking sequence as hexagonal (*h*) or cubic (*k*) configuration.

(b) Schematic side-view of 6H−SiC (0001) projected in (11$\bar{2}$0) plane with the six possible surface terraces S1*, S2*, S3*, S1, S2, S3. The surfaces Sn and Sn* (n = 1-3) are equivalent but rotated by 60° related to each other. The terrace widths are strongly reduced. A terrace width of ∼240 nm is estimated for a miscut angle of 0.06° consistent with experimental results. The arrows mark the different retraction velocities of the Si-C layers in the step retraction model, which are related to the individual surface layer hexagonality.

(c) Basic terrace step patterns of the 6H-SiC surface developing in the step retraction model. In the initial state, the individual S1, S2, and S3 terraces are separated by single SiC monolayer steps (1L) of ∼0.25 nm in height. In the intermediate state, the S1 surface with the fastest retraction velocity has disappeared, and an alternating sequence of S2 and S3 surfaces remain with steps of ∼0.25 nm (1L) above S2 and ∼0.5 nm (2L) above S3. After advanced step retraction S2 becomes wider than the S3 terrace since the step velocity of S3 is faster than S2, which is depicted here. In the final state, the most stable S2 terraces remain with ∼0.75 nm (3ML) in height.

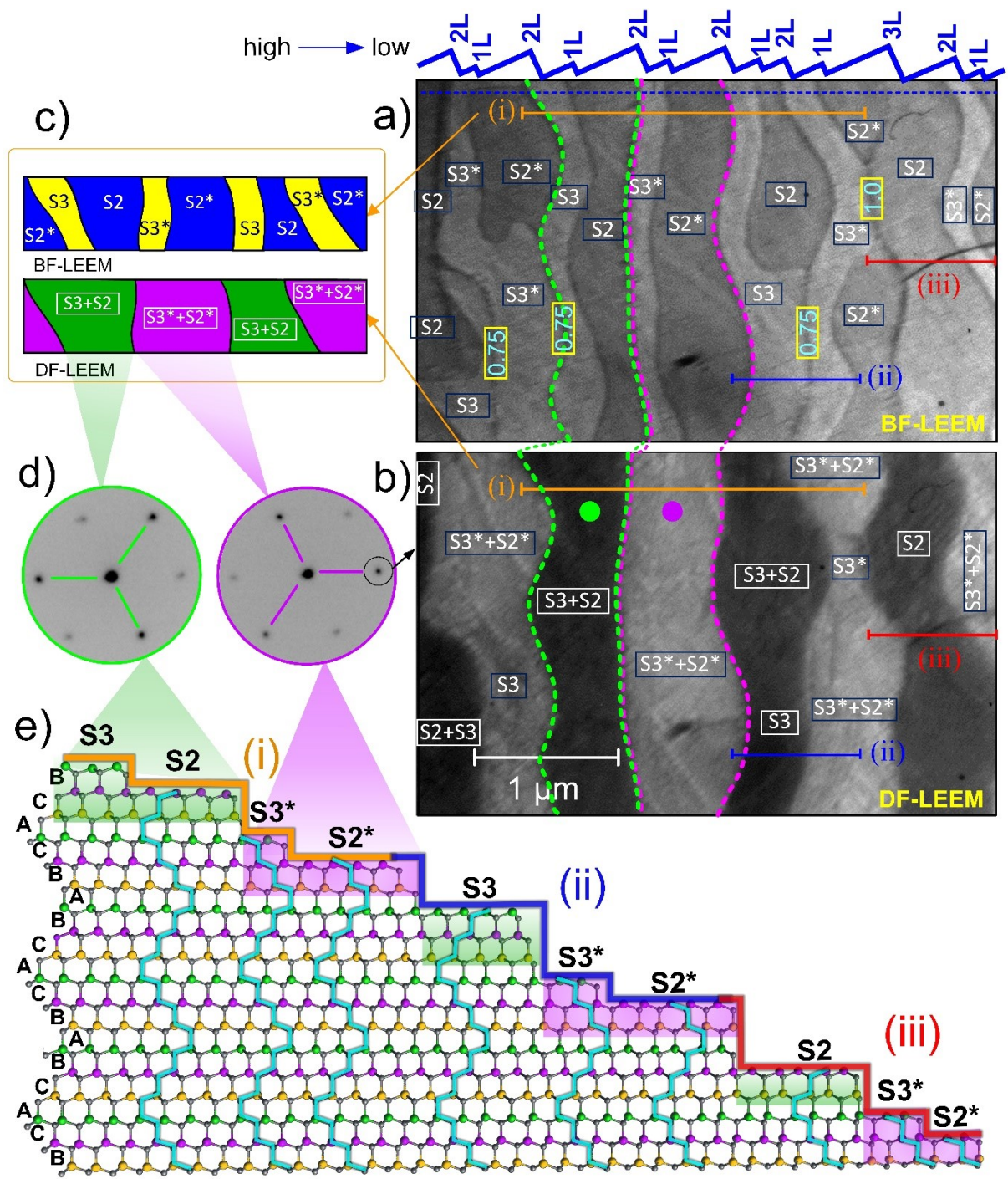

**Figure 3.**

(a) BF- LEEM ($E = 4.4$ eV) image of PASG monolayer graphene on 6H-SiC(0001) showing a stacking-related reflectivity contrast. The terraces can be distinguished by their reflectivity and are labeled S2, S2*, S3, S3*, as explained in the text. From this, a height profile (top blue line) is deduced for the upper part of the LEEM image (along the dotted blue line).

(b) Dark-field LEEM ($E = 11$ eV) image of the same surface area using the diffraction spot marked with a black circle in (d). The contrast is caused by the 60° crystal rotation of the terminating SiC layers of the substrate. Areas of the same brightness are labeled by the SiC terraces as deduced from the BF-LEEM image.

(c) Sketch of the BF- and DF-LEEM image for a selected area (orange line (i)) in (a) and (b), which represents a region with a regular 1L/2L step pattern.

(d) Two μ-LEED ($E = 37$ eV) patterns from neighboring areas with different dark-field LEEM contrast (marked by the green and violet dot in (b)) The 60° rotation of the satellite spots indicates the corresponding SiC crystal rotation.

(e) Schematic step restructuring model of the 6H-SiC substrate during epi-graphene growth for three typical step patterns observed in the LEEM images (a) and (b). Buffer and graphene layers are not shown for clarity. Area (i) demonstrates the characteristic regular pattern of 1L/2L step pairs with the S3/S2 and S3*/S2* terrace sequence along line (i) in both LEEM images. The cyan-colored line-tracks indicate the different SiC crystal rotation below the terrace pairs S3/S2 and S3*/S2*, which give rise to the DF-LEEM contrast. Areas (ii) and (iii) with an irregular step sequence including ~0.75 nm (3L) steps fully explain the observed BF-and DF-LEEM contrast patterns along the line (ii) and (iii) in (a) and (b).

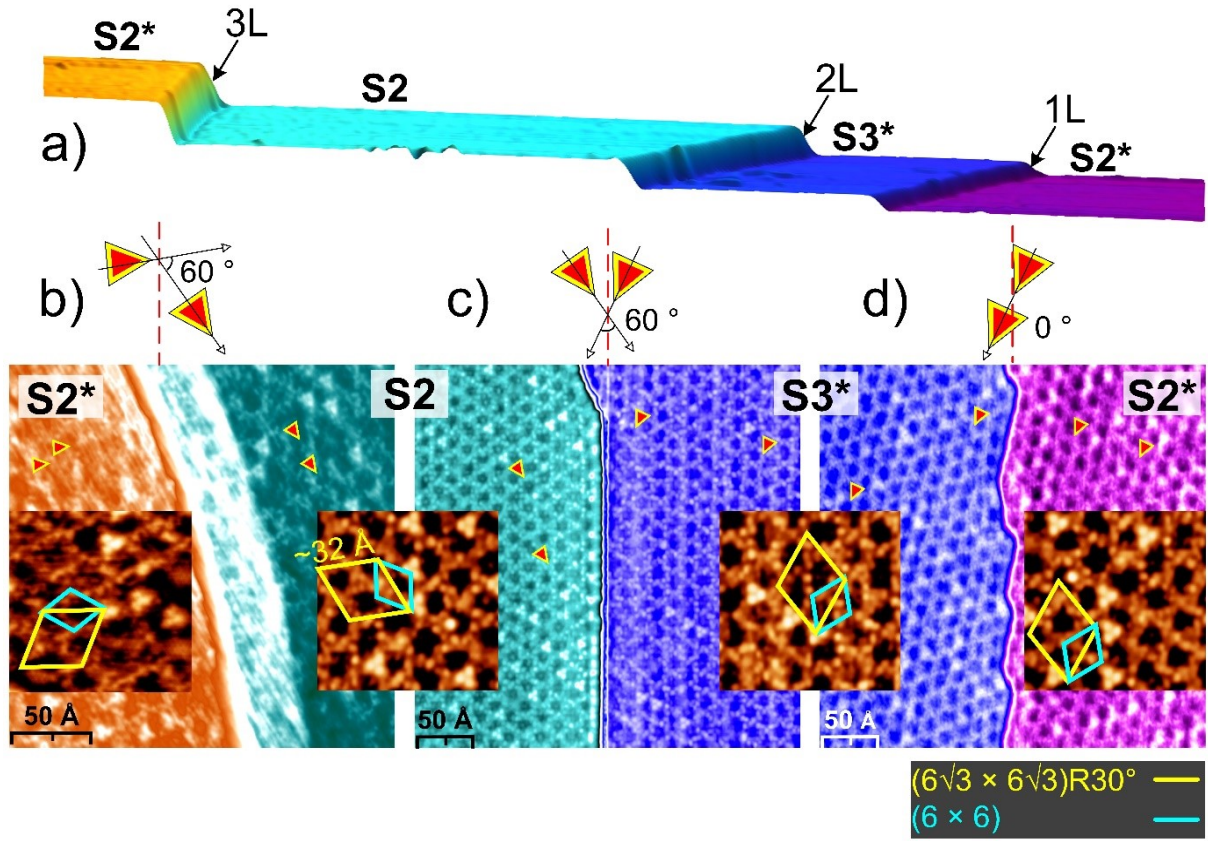

## Figure 4.

(a) Scanning tunneling microscopy image of PASG monolayer graphene on 6H-SiC with non-identical surface terminations measured in the area along the line type (iii) in the LEEM image of **Figure 3**a. The assignment of the surfaces (S2*, S2, S3*, and S2*) and the corresponding step heights are a direct result of the interpretation of the LEED images within the step retraction model and are schematically displayed in the area (iii) of **Figure 3**e. The images were recorded at 77 K with an Omicron low-temperature STM using a tungsten tip in the constant-current STM mode at 0.4 nA and 1.7 V.

Atomic resolution STM images of the graphene buffer layer on neighboring terraces around the 3L, 2L, and 1L step edges, respectively, shown in (a). (b) shows the transition area from terrace S2* to S2, which is correlated with a 60° rotation of the SiC substrate, see the area (iii) in **Figure 3**a and e. This rotation manifests itself in a 60° rotation of triangular-shaped structures partly marked by the red/yellow triangles. For clarity, the directions of the triangles are sketched above each image. These triangular structures span a (6 × 6) nanomesh, which also rotates by 60° indicated by the blue diamond in the high-resolution insets. The buffer layer characteristic $(6\sqrt{3} × 6\sqrt{3})R30°$ super-lattice is indicated by a yellow diamond and it follows the 60° rotation of the SiC surface. (c) The transition from an S2 to an S3* terrace is also characterized by a 60° rotation of the triangular structures, the (6 × 6) nanomesh, and the buffer layer superlattice. (d) For the S3* and S2* transition, no rotation of the triangles and, therefore, of the buffer layer is observed in agreement with the missing SiC crystal rotation.

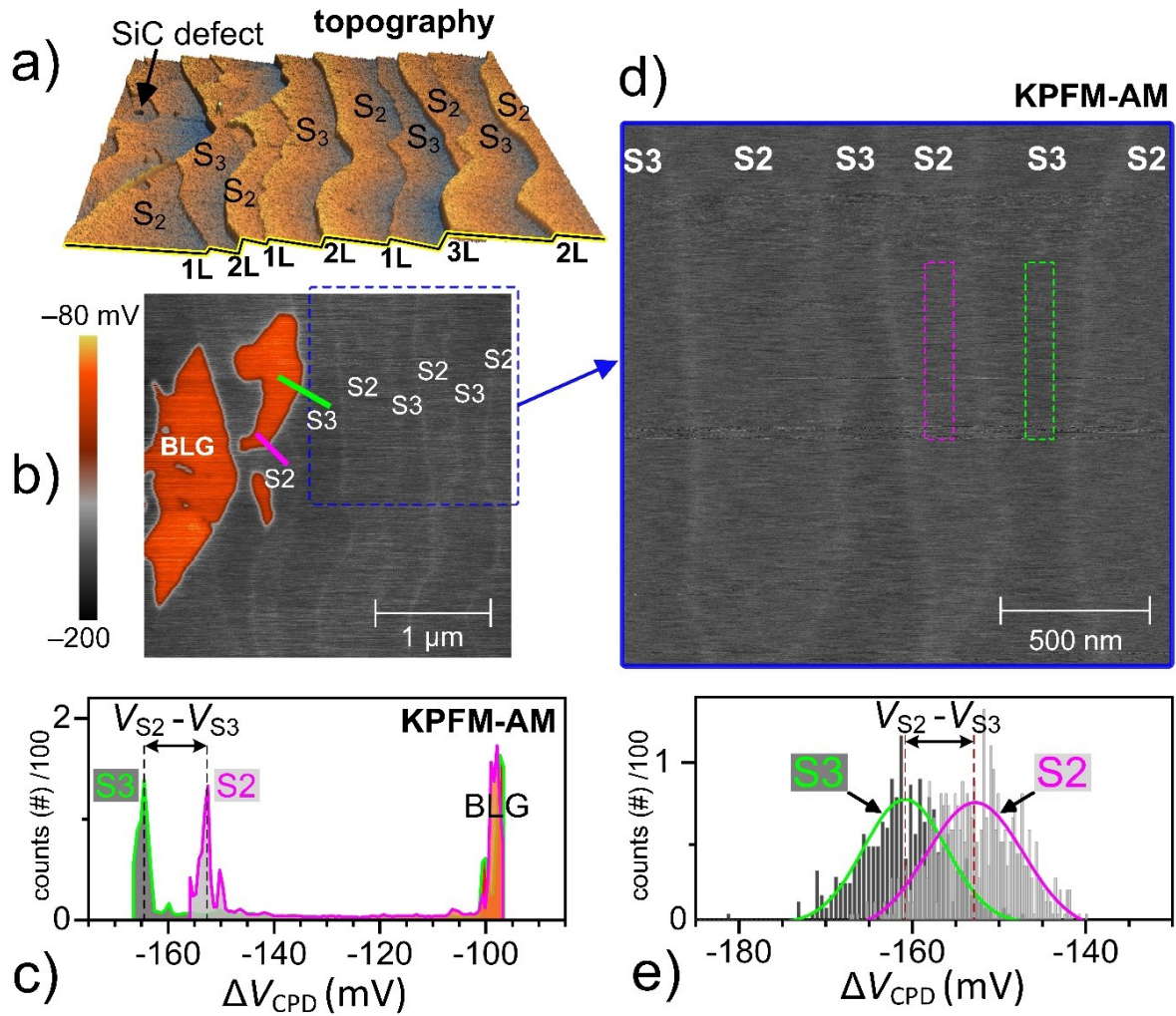

**Figure 5.**

Quantitative measurements of surface potential and work function by Kelvin-Probe Force-Microscopy (KPFM) in amplitude mode (AM) of binary 1L/2L stepped monolayer graphene on Si-terminated 6H-SiC.

(a) The AFM topography image allows an unambiguous assignment of the surface terraces S2 and S3 based on the step height sequence, as explained in the text.

(b) Surface potential map from KPFM-AM measurement of the same surface area. The red areas mark bilayer graphene (BLG) spots, which have formed at a SiC defect. Marked are the positions of two line scans (green and magenta-colored lines) correlated with an S3 and an S2 terrace, respectively, for the histogram evaluation.

(c) Histogram of the surface potential differences along the two line scans S2 (magenta) and S3 (green) indicated in the inset of (b). A clear difference of the potential values for the S2 and S3 surface is observed with $V_{S2} - V_{S3} = 12 \pm 2$ mV. A separation of ~60 mV to the potential values of the BLG is clearly visible.

(d) Section of KPFM image in (b) (blue square) showing the potential contrast of monolayer graphene on the terraces S2 and S3. The green and magenta-colored rectangles ($100 \times 600$ nm$^2$) indicate the area for the extraction of the potential values shown in (e).

(e) Histogram of the surface potential difference (median values) extracted from the green (S3 terrace) and magenta (S2 terrace) rectangles shown in (d). A potential difference $V_{S2} - V_{S3} = 9 \pm 2$ mV is estimated.

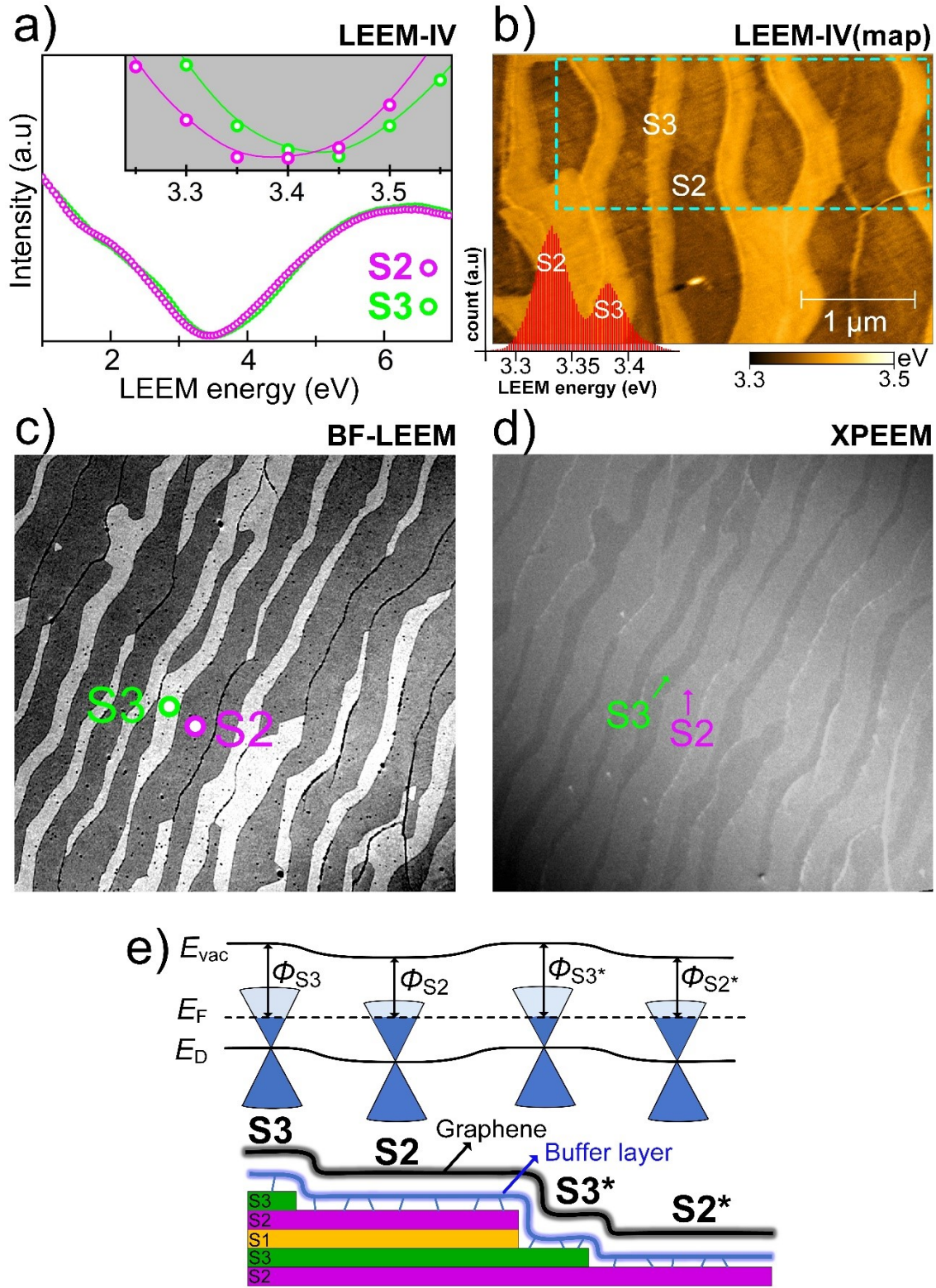

**Figure 6.**

(a) Two low-energy electron spectra (LEEM-IV) spectra taken at two different terraces S2 and S3, as indicated in (c). The inset shows the minimum position and a difference in energy of ~60 meV between the S2 and S3 curve.

(b) The LEEM-IV map shows the lateral distribution of the minimum energy from the LEEM-IV spectra in an area of 4 μm × 3 μm. This LEEM-IV map is taken from the same area as the BF-LEEM image in **Figure 3**a, and the comparison reveals a congruent contrast pattern related to the SiC terraces underneath. The histogram (inset) shows the distribution of LEEM-IV minimum energies taken from the area marked with the dashed-cyan rectangle. The histogram clearly indicates an energy difference of 60 $\pm$10 meV between graphene on the S2 and S3 terminations.

(c) BF-LEEM image from another 1L/2L stepped monolayer graphene sample. The indicated dots S2 and S3 mark the position where the LEEM-IV spectra in (a) were measured. The incident electron energy was 2.7 eV, and the field of view (FoV) is 10 µm.

(d) X-ray photoemission electron microscopy (XPEEM) image taken at the same positions as the BF-LEEM image in (c). The field of view (FoV) is 10 µm, X-ray excitation of 80 eV, and detection at 1 eV. Note the inverse contrast. The XPEEM contrast is related to the work function of the graphene. The higher work function terraces show a darker contrast. The white lines stem from 3L step edges between terraces of the same SiC surface termination.

(e) Schematic energy diagram of epitaxial monolayer graphene on the 6H-SiC terraces S2 and S3 as derived from XPEEM and KPFM measurements. The variation of the work functions $\phi$ at S2 and S3 terraces indicates a shift of the Dirac cones by the same amount (Dirac energy $E_D$). This results in a spatial modulation of the graphene surface potential.